\documentclass[10pt,a4paper]{article}
\usepackage{jheppub}
\usepackage[utf8]{inputenc}
\usepackage[english]{babel}
\usepackage{amsmath}
\usepackage{amsfonts}
\usepackage{amssymb}

\newcommand{\bz}{\boldsymbol z }
\newcommand{\bfi}{\boldsymbol \varphi }

\title{A novel derivation of the Marchenko-Pastur law through analog bipartite spin-glasses\footnote{The Authors are pleased to dedicate this paper to Giorgio Parisi in occasion of his seventieth birthday.}}
\author[a,b]{Elena Agliari,}
\author[c,d]{Francesco Alemanno,}
\author[c,d,e]{Adriano Barra,}
\author[c,d,e]{Alberto Fachechi}
\affiliation[a]{Dipartimento di Matematica, Sapienza Universit\`a di Roma, Italy}
\affiliation[b]{GNFM-INdAM Sezione di Roma, Italy}
\affiliation[c]{Dipartimento di Matematica e Fisica Ennio De Giorgi, Universit\`a del Salento, Italy}
\affiliation[d]{GNFM-INdAM Sezione di Lecce, Italy}
\affiliation[e]{INFN, Istituto Nazionale di Fisica Nucleare, Sezione di Lecce, Italy}
\emailAdd{elena.agliari@uniroma1.it}
\emailAdd{francesco.alemanno@le.infn.it}
\emailAdd{adriano.barra@unisalento.it}
\emailAdd{alberto.fachechi@le.infn.it}
\abstract{In the last decades, statistical mechanics of disordered systems (mainly spin glasses) has become one of the main tool to investigate complex systems, probably due to the celebrated Replica Symmetry Breaking scheme of Parisi Theory and its deep implications.
\newline
In this work we consider the {\em analog bipartite spin-glass} (or {\em real-valued restricted Boltzmann machine} in a neural network jargon), whose variables (those quenched as well as those dynamical) share standard Gaussian distributions. First, via Guerra's interpolation technique, we express its quenched free energy in terms of the natural order parameters of the theory (namely the self- and two-replica overlaps), then, we re-obtain the same result by using the replica-trick: a mandatory tribute, given the special occasion.
Next, we show that the quenched free energy of this model is the functional generator of the moments of the correlation matrix among the weights connecting the two layers of the spin-glass ({\it i.e.}, the {\em Wishart matrix} in random matrix theory or the {\em Hebbian coupling} in neural networks): as weights are quenched stochastic variables, this plays as a novel tool to inspect random matrices. In particular, we find that the Stieltjes transform of the spectral density of the correlation matrix is determined by the (replica-symmetric) quenched free energy of the bipartite spin-glass model. In this setup, we re-obtain the Marchenko-Pastur law in a very simple way.}

\keywords{Spin Glasses, Disordered Systems, Statistical Mechanics, Random Matrix Theory}
\begin{document}
\maketitle

\section{Introduction}

It did not take long for scientists to realize the impressive representational power of the Sherrington-Kirkpatrick model ({\it i.e.}, the {\em simplest} mean-field spin glass), especially in relation to its low temperature behaviour with the spontaneous hierarchical organization of its thermodynamical states, as painted by Parisi at the turn of the seventies and eighties \cite{Giorgio1,Giorgio2,Giorgio3} (and duly mathematically confirmed in recent times by Guerra \cite{Guerra-Broken}, Panchencko \cite{Panchenko} and Talagrand \cite{talaProof}). Since then, spin glasses quickly became the {\em harmonic oscillators} of complex systems, with applications ranging from Biology to Engineering. As for the former, the spin-glass framework applies over different scales, from the extracellular world of neural \cite{Albert2,Giorgio-Neural} and  immune \cite{Agliari-Immune,Giorgio-Immune} networks to the inner world of protein folding \cite{protein} and gene regulatory networks \cite{gene} and even questioning about evolution \cite{Kaufmann,Peliti}; as for the latter, spin-glasses are extensively exploited in machine learning \cite{Coolen,Giorgio-Learning} and computer science \cite{Zecchina1,Zecchina2}, tackling computational complexity from an entirely statistical-mechanics driven perspective \cite{CompCompl,Nishimori} (of course, this is far from being an exhaustive list). Further, spin-glass theory gave several hints even in more theoretical contexts (quoting Talagrand, ``spin glasses are heaven for mathematicians'' \cite{Tala}) as their studies inspired several fields of Mathematics such as random matrix theory  \cite{RMT}, variational calculus \cite{Auffinger}, probability theory \cite{Bovierbook}, PDE theory \cite{dynamic} and much more.
\newline
In this paper we investigate a hidden link between bipartite spin-glasses and the Marchenko-Pastur law.
To this goal we consider a {\em fully analog} bipartite system, whose quenched and dynamical variables ({\it i.e.}, the couplings and the spins, respectively) are all drawn from a normal distribution $\mathcal{N}[0,1]$, and we look for an explicit expression of its quenched free energy: this is achieved by means of Guerra's interpolating technique and, independently, by means of the celebrated replica trick. Once the resulting free energy is extremized over the order parameters, en route for their self-consistent expression, we obtain the latter in terms of an algebraic set of equations, whose solution captures the momenta of the Wishart matrix stemming from the weights among the two parties.
\newline
More explicitely, if the two parties have sizes $N$ and $P=\alpha N$ (with $\alpha \in \mathbb{R}^+$), we can define their mutual connections via a weight matrix $\xi_i^{\mu}$ ($i \in (1,...,N)$ and $\mu \in (1,...,P)$), such that the weight correlation matrix reads as $C_{\mu \nu}= N^{-1}\sum_{i}^{N}\xi_i^{\mu}\xi_j^{\mu}$. Notice that, by definition, the matrix $\bold{C}$ is, according to the perspective, a Wishart matrix as well as a Hebbian kernel \cite{Amit,Coolen} and its spectrum obeys the well-known Marchenko-Pastur law. Then, we prove that the quenched free energy of the fully analog bipartite spin glass is the generating functional of the moments of such a correlation matrix. This link allows the usage of the Stieltjes-Perron inversion formula to reconstruct the spectral density of the correlation matrix, thus obtaining, from a pure statistical mechanics framework, the Marchenko-Pastur distribution.
\newline
Extensive comparison between our theory and simulations shows full agreement between the analytical and the numerical sides of the present investigation.
\newline
As a final remark we stress that another powerful link between Disordered Statistical Mechanics and Random Matrix Theory was settled  in \cite{Kuhn} where the Wigner semicircle distribution and the Marchenko-Pastur law  were re-obtained by an extensive use of the cavity field technique, another primary technique in spin-glasses \cite{Barra-JSP2006,GuerraCavity,MP0,MPV}, but directly applied in a pure random matrix context.

\section{The analog bi-partite spin-glass}

We start by introducing the model, described by the following energy {\em cost function}
\begin{equation}\label{model}
E(\bz,\bfi \vert \xi):=-\frac{1}{\sqrt{N}}\sum_{\mu, i}^{P,N}\xi^\mu _i z_\mu \varphi_i.
\end{equation}
This is a bipartite system composed by two different layers of standard Gaussian variables $z_\mu$ and $\varphi_i$  with symmetric interactions $\xi_i^{\mu}$, whose strength is also sampled i.i.d. from $\mathcal{N}[0,1]$. The indices $i$ and $\mu$ run respectively in the ranges $1,\dots, N$ and $1,\dots,P$.\footnote{We point out that this model could be seen as a fully analog random restricted Boltzmann machine (RBM) \cite{BM1,Barra-RBMsPriors1,Barra-RBMsPriors2,Monasson}.} We will consider the so-called {\em high-storage} case where $P =\alpha N$, with $\alpha \in \mathbb{R}^+$, for large $N$, namely the challenging regime to explore when studying bipartite spin glasses \cite{bipartiti,how-glassy} or restricted Boltzmann machines \cite{Agliari-Dantoni,Coolen} (and the related Hopfield model \cite{Agliari-PRL1,Albert1,Albert2,Mezard,Monasson}).
\newline
The partition function for such a system is therefore defined as
\begin{equation} \label{eq:part}
	Z_{N,P}(\beta|\xi)  := \int d\mu (\bz,\bfi)\exp\Big(\frac{\beta}{\sqrt{N}}\sum_{\mu, i}^{P, N}\xi^\mu _i z_\mu \varphi_i\Big),
\end{equation}
where $\beta \in \mathbb{R}^+$ tunes the level of noise and
\begin{equation}
 d\mu (\bz,\bfi):= \prod_{\mu, i}^{P, N}\frac{dz_\mu}{\sqrt{2\pi}}\exp(-z_\mu^2/2)\frac{d\varphi_i}{\sqrt{2\pi}}\exp(-\varphi_i^2/2)
\end{equation}
is the multidimensional Gaussian measure. The key quantity to study in order to have a picture of the properties of the model is its quenched free-energy defined as
\begin{equation}
	\mathit f(\alpha,\beta) := \lim_{N \to \infty} \frac{1}{N}\mathbb E_\xi \log\int d\mu (\bz,\bfi)\exp\Big(\frac{\beta}{\sqrt{N}}\sum_{\mu, i}^{P,N} \xi^\mu _i z_\mu \varphi_i\Big),
\end{equation}
where $\mathbb E_\xi$ is the average over all possible weight (or {\em coupling} or  {\em pattern}) realizations, {\it i.e.}
$$
\mathbb E_x f(x)=  \int dx f(x) e^{-x^2/2}.
$$

\section{Replica symmetric solution of the quenched free energy}
\subsection{The interpolation scheme}
The computation of the quenched free energy in the thermodynamic limit can be performed by means of a Guerra's interpolation method (whose genesis lies in the treatment of the analog Hopfield model \cite{Barra-JSP2010}). More precisely, once introduced an interpolating parameter $s \in (0,1)$, the scalars $A,B,C,D$ (whose explicit values will be set {\it a fortiori}), $N$ i.i.d. Gaussian fields $\eta_i \sim \mathcal{N}[0,1]$, $i \in (1,...,N)$ and $P$ i.i.d. Gaussian fields $\psi_{\mu} \sim \mathcal{N}[0,1]$, $\mu \in (1,...,P)$, the interpolating free energy is defined as
\begin{equation}\label{interpol}
\begin{split}
	\mathit f (s):=\frac{1}{N}\mathbb E_{\xi,\eta,\psi} \log\int d\mu (\bz,\bfi)\exp&\Big[\sqrt{s}\frac{\beta}{\sqrt{N}}\sum_{\mu i}\xi^\mu _i z_\mu \varphi_i\\&
	+\sqrt{1-s}	(A\sum_i \eta_i \varphi_i+B\sum_{\mu}\psi_\mu z_\mu)\\&
	+\frac{1-s}{2}(C\sum_{\mu}z_\mu^2+D\sum_i \varphi_i^2)
	\Big].
\end{split}
\end{equation}
In the first line, we introduced the contribution coming from the model under consideration (which is of course reproduced for $s=1$, {\it i.e.}, in the thermodynamic limit $f(s=1)=f(\alpha,\beta)$); the second line is an effective mean-field contribution, statistically representing the {\em external field} produced by each layer on the other one (but where we replaced the original two-body interactions with a one-body coupling); the third line contains terms necessary to properly rescale the second moments of the dynamical variables $(\bz,\bfi)$.
\newline
Note that now the average $\mathbb E_{\xi}$ over quenched variables has to be replaced by $\mathbb E_{\xi,\eta,\psi} \equiv \mathbb E$ such that the average is performed over all the quenched random variables, as  stressed by the subscript.
\newline
In order to compute the free energy of the system \eqref{eq:part}, we adopt the elementary sum rule
\begin{equation}\label{sum-rule}
	\mathit f(s=1)= \mathit f(s=0)+\int _0^1 \frac{df(s)}{ds} ds .
\end{equation}
The computation of the derivative in (\ref{sum-rule}) is straightforward:
\begin{equation}
\begin{split}
 \frac{df(s)}{ds} =\frac{1}{N}\mathbb{E}\Bigg[&\frac{1}{2\sqrt s}\frac{\beta}{\sqrt N}\sum_{i,\mu=1}^{N,P} \xi^\mu_i \langle z_\mu \varphi_i\rangle-\frac{1}{2 \sqrt{1-s}}	\left (A\sum_{i=1}^N \eta_i \langle\varphi_i\rangle+B\sum_{\mu=1}^P\psi_\mu \langle z_\mu\rangle \right)\\&
-\frac{1}{2} \left (C\sum_{\mu=1}^P \langle z_\mu^2\rangle+D\sum_{i=1}^N \langle\varphi_i^2\rangle \right)
\Bigg].
\end{split}
\end{equation}
By simple applications of the Wick theorem [$\mathbb{E}_x xF(x)= \mathbb{E}_x \partial_xF(x)$] over the Gaussian variables, we have
\begin{equation}
\begin{split}
\mathbb E \,\xi^\mu_i \langle z_\mu \varphi_i\rangle&=\sqrt{s/N}\beta ~ \mathbb E \,(\langle z_\mu^2 \varphi_i ^2\rangle - \langle z_\mu \varphi_i\rangle^2),\\
\mathbb E \, \eta_i \langle \varphi _i \rangle&= A\sqrt{1-s} ~ \mathbb E \,(\langle \varphi_i^2\rangle- \langle \varphi_i\rangle^2),\\
\mathbb E \, \psi_\mu \langle z_\mu \rangle&= B\sqrt{1-s} ~ \mathbb E \,(\langle z_\mu^2\rangle- \langle z_\mu\rangle^2).
\end{split}
\end{equation}
Then,
\begin{equation}\label{streaming}
\begin{split}
\frac{df(s)}{ds}=&\frac{1}{N}\mathbb{E}\Big[ \frac{\beta^2}{2N}\sum_{\mu, i=1}^{P,N}(\langle z_\mu^2 \varphi_i^2\rangle-\langle z_\mu \varphi_i \rangle^2)-
\frac{A^2}{2}\sum_{i=1}^N (\langle \varphi_i^2\rangle - \langle \varphi_i \rangle^2)-\frac{B^2}{2}\sum_{\mu=1}^P(\langle z_\mu^2\rangle -\langle z_\mu \rangle^2)\\&
-\frac{1}{2}(C\sum_{\mu=1}^P\langle z_\mu^2\rangle+D\sum_{i=1}^N \langle\varphi_i^2\rangle)
\Big]
\end{split}
\end{equation}
We now introduce the self- and the two-replica overlaps within each layer, respectively as
\begin{equation}
\begin{split}
q_{11}&= \frac{1}{N}\sum_{i=1}^N \varphi_i ^2,\qquad\qquad p_{11}= \frac1P \sum_{\mu=1}^P  z_\mu^2,\\
q_{12}&= \frac{1}{N}\sum_{i=1}^N \varphi_i ^{(1 )}\varphi_i ^{(2 )},\qquad p_{12}= \frac1P \sum_{\mu=1}^P z_\mu^{(1 )} z_\mu^{(2 )},
\end{split}
\end{equation}
where the superscript $(1),(2)$ are replica indices, and we use these overlaps to recast the r.h.s. of eq. (\ref{streaming}) as
\begin{equation}
\begin{split}
\frac{df(s)}{ds} = \frac{\alpha\beta^2}{2}\mathbb E\Big[&\langle p_{11}q_{11}\rangle -\frac{A^2+D}{\alpha\beta^2}\langle q_{11}\rangle -\frac{B^2+C}{\beta^2}\langle p_{11}\rangle-( \langle p_{12}q_{12}\rangle -\frac{A^2}{\alpha\beta^2}\langle q_{12}\rangle -\frac{B^2}{\beta^2}\langle p_{12}\rangle)\Big].
\end{split}
\end{equation}
Finally, by choosing
\begin{equation} \label{para}
A=\beta \sqrt{ q_N}, \quad B =\beta\sqrt{ p_N},\quad C=\beta^2 ( q_D- q_N),\quad D=\alpha\beta^2 ( p_D- p_N),
\end{equation}
we can rewrite the streaming of the interpolating free energy as
\begin{equation} \label{eq:stream}
\frac{df(s)}{ds}  =\frac{\alpha\beta^2}{2}\mathbb E\Big[\langle (p_{11}- p_D)(q_{11}- q_D)\rangle- p_D  q_D-(\langle (p_{12}- p_N)(q_{12}- q_N)\rangle- p_N  q_N)\Big].
\end{equation}
The crucial point now is to fix $q_{D,N}$ and $p_{D,N}$. We notice that, in a replica-symmetry regime,\footnote{We stress that, for these analog models, RSB is not expected and, at least for the single-layer Gaussian spin glass, the replica symmetric free energy is correct at all the values of the noise \cite{gaussian-spinglass,BenArous}.} the overlaps converge to their average values (meaning that fluctuations are suppressed) in the thermodynamic limit. Therefore, if we interpret $ q_{D,N}$ and $ p_{D,N}$ as respectively the (self-averaging) equilibrium values of the diagonal and non-diagonal overlaps in the thermodynamic limit, we simply have $d_s f(s) = \tfrac{\alpha\beta^2}{2} p_N  q_N-\tfrac{\alpha\beta^2}{2} p_D  q_D$, that is clearly $s$-independent. Consequently, the integral in the sum rule (\ref{sum-rule}) is trivial as it coincides with the multiplication by one.
\par\medskip
The one-body term ({\it i.e.}, $f(s=0)$) is easy to evaluate, since it is a one-body calculation, returning
\begin{equation} \label{eq:zero}
\begin{split}
f(s=0)&= \frac{1}{N}\mathbb E \ln \int d\mu (\bz,\bfi)\exp\Big(A\sum_i \eta_i \varphi_i +B\sum_{\mu }\psi_\mu z_\mu+C\sum_{\mu }z_\mu^2 +D\sum_i \varphi_i^2\Big)=\\
&=-\frac{\alpha}{2}\log(1-C)-\frac12 \log (1-D)+\frac{A^2}{2(1-D)}+\frac{\alpha B^2}{2(1-C)}.
\end{split}
\end{equation}
By combining (\ref{sum-rule}), (\ref{eq:stream}) and (\ref{eq:zero}), we get the quenched free energy related to the analog bipartite spin-glass as
\begin{equation}\label{eq:freestoc}
\begin{split}
f(\alpha,\beta)=&-\frac{\alpha}{2}\log[1-\beta^2( q_D- q_N)]-\frac{1}{2}\log[1-\alpha \beta^2( p_D- p_N)]\\&+\frac{\alpha\beta^2  p_N}{2[1-\alpha\beta^2( p_D - p_N)]}+\frac{\alpha\beta^2  q_N}{2[1-\beta^2( q_D - q_N)]}+\frac{\alpha\beta^2}{2} p_N  q_N-\frac{\alpha\beta^2}{2} p_D  q_D.
\end{split}
\end{equation}

\subsection{The replica trick route}

To conclude this investigation, recalling that this is a dedicated contribution to celebrate the seventieth birthday of Giorgio Parisi, it seems mandatory to re-obtain the previous results with the first route paved in order to embed replica symmetry breaking into a mathematical framework: the replica trick \cite{MPV}.\footnote{The problem still consists in obtaining an explicit expression, in terms of the self and two-replica overlaps, of the quenched free energy of the model (\ref{model}): indeed we have chosen to study and dedicate this model to {\em Giorgio70} also because - in the present context - the two routes (interpolation and replica trick) are quite transparent in all their steps of usage such that it is possible to appreciate the evolution of these mathematical weapons developed by the various Schools along the decades.} In the original replica-trick framework \cite{MPV}, the quenched free energy is represented as
\begin{eqnarray}
f(\alpha,\beta) = \lim_{N\rightarrow\infty}\lim _{n\rightarrow 0}\frac{\mathbb E Z_{N,P}^n(\beta|\xi) -1}{n N}.
\end{eqnarray}
The $n$-th power of the partition function is therefore naturally expressed in terms of replicas as
\begin{eqnarray}
Z_{N,P}^n(\beta|\xi) =\prod _{\gamma =1}^n Z_{N,P}^{(\gamma)}(\beta|\xi)= \int \left(\prod _{\gamma=1}^n d\mu (\bz^{\gamma},\bfi ^{\gamma}) \right) \exp\Big(\frac{\beta}{\sqrt N}\sum_{i, \mu, \gamma=1}^{N,P,n}\xi^\mu _i z_\mu ^\gamma \varphi^\gamma _i\Big).
\end{eqnarray}
In this way, one can directly perform the average over the quenched noise (the couplings $\boldsymbol{\xi}$), so that
\begin{equation}
\mathbb E Z_{N,P}^n(\beta|\xi) =\int \left(\prod _\gamma d\mu (\bz^{\gamma},\bfi ^{\gamma})\right) \exp\Big(\frac{\beta^2}{2 N}\sum_{i, \mu }\sum_{\gamma, \delta} z_\mu ^\gamma z_\mu ^\delta \varphi^\gamma _i \varphi ^\delta _i\Big).
\end{equation}
In order to linearize the problem in the dynamical variables, one can introduce the overlaps in each layer by inserting, for each kind of overlap and for each couple of replicas, a Dirac delta, namely
\begin{equation}
\delta (q_{\gamma\delta}-N^{-1}\sum_i \varphi_i ^\gamma \varphi_i ^\delta) \delta (p_{\gamma\delta}-P^{-1}\sum_\mu z_\mu^\gamma z_\mu ^\delta),
\end{equation}
then, by using the Fourier representation of the delta function, we can rewrite the partition function as
\begin{equation}
\begin{split}
\mathbb E Z_{N,P}^n (\beta|\xi) & =\int \left(\prod _\gamma d\mu (\bz^{\gamma},\bfi ^{\gamma}) \right) \left(\prod_{\gamma,\delta}dq_{\gamma \delta}dp_{\gamma \delta} \frac{N d\bar q_{\gamma \delta}}{2\pi} \frac{P d\bar q_{\gamma \delta}}{2\pi} \right)\times \\&
\times \exp\Big(iN\sum_{\gamma, \delta}\bar q_{\gamma \delta}(q_{\gamma \delta}-\tfrac{1}{N}\sum_i \varphi_i ^\gamma \varphi_i ^\delta)+iP\sum_{\gamma, \delta}\bar p_{\gamma \delta}(p_{\gamma \delta}-\tfrac{1}{P}\sum_{\mu} z_\mu^\gamma z_\mu ^\delta)+\frac{\beta^2 P}{2}\sum_{\gamma, \delta}p_{\gamma \delta}q_{\gamma \delta} \Big),
\end{split}
\end{equation}
where $\bar q_{\gamma \delta}$ and $\bar p_{\gamma \delta}$ are, respectively, the conjugates of $q_{\gamma \delta}$ and $p_{\gamma \delta}$, for any $\gamma, \delta=1,..., n$. At this point, we can easily perform the Gaussian integration over the dynamical variables, yielding (after a trivial rescaling of the conjugates $\bar q_{\gamma \delta} \rightarrow 2^{-1} i \bar q_{\gamma \delta}$ and $\bar p_{\gamma \delta} \rightarrow 2^{-1}i \bar p_{\gamma \delta}$)
\begin{equation}
\begin{split}
\mathbb E Z^n_{N,P}(\beta|\xi) & =\int \left(\prod_{\gamma\delta}dq_{\gamma\delta}dp_{\gamma\delta} \frac{iN d\bar q_{\gamma\delta}}{4\pi} \frac{iP d\bar q_{\gamma\delta}}{4\pi} \right)\times \\&
\times \exp\Big(-\frac N2\sum_{\gamma\delta}\bar q_{\gamma\delta}q_{\gamma\delta}-\frac{\alpha N}2\sum_{\gamma\delta}\bar p_{\gamma\delta}p_{\gamma\delta}+\frac{\alpha\beta^2 N}{2}\sum_{\gamma\delta}p_{\gamma\delta}q_{\gamma\delta}\\&-\frac{\alpha N}{2}\log \det(1-\bar {\boldsymbol p}) -\frac{ N}{2}\log \det(1-\bar {\boldsymbol q}) \Big),
\end{split}
\end{equation}
where $\bar {\boldsymbol q}$ and $\bar {\boldsymbol p}$ are the conjugate overlap matrices. By taking the intensive logarithm of the expression above, we get the following quenched free energy 
\begin{equation}
\begin{split}
f(\alpha,\beta) &=-\frac{1}{2n}\sum_{\gamma\delta}\bar q_{\gamma\delta}q_{\gamma\delta}-\frac{\alpha }{2n}\sum_{\gamma\delta}\bar p_{\gamma\delta}p_{\gamma\delta}+\frac{\alpha\beta^2 }{2n}\sum_{\gamma\delta}p_{\gamma\delta}q_{\gamma\delta}\\&-\frac{\alpha }{2n}\log \det(1-\bar {\boldsymbol p}) -\frac{ 1}{2n}\log \det(1-\bar {\boldsymbol q}).
\end{split}
\end{equation}
The replica symmetry ansatz, in this context, reads as
\begin{equation}
\begin{split}
q_{\gamma\delta}=&q_D \delta_{\gamma\delta}+q_N (1-\delta_{\gamma\delta}),\\
p_{\gamma\delta}=&p_D \delta_{\gamma\delta}+p_N (1-\delta_{\gamma\delta}),\\
\bar q_{\gamma\delta}=&\bar q_D \delta_{\gamma\delta}+\bar q_N (1-\delta_{\gamma\delta}),\\
\bar p_{\gamma\delta}=&\bar p_D \delta_{\gamma\delta}+\bar p_N (1-\delta_{\gamma\delta}).
\end{split}
\end{equation}
Then, by straightforward computations, assuming the commutativity of the infinite volume limit and  the zero-replica limit,\footnote{A nice note could be added here: in the occasion of David Sherrington's 70th birthday, the contribution \cite{Mingione} proved that, at least for the Sherrington-Kirkpatrick spin glass, the infinite volume limit and the zero replicas limit do commute.} we have
\begin{equation}\label{eq:frep}
\begin{split}
f(\alpha,\beta) &=\frac{\alpha\beta^2}{2}(p_D q_D - p_N q_N)-\frac{\alpha}{2}(\bar p_D p_D - \bar p_N p_N)-\frac12 (\bar q_D q_D -\bar q_N q_N)\\&
-\frac{\alpha}{2}\Big(\log [1-\bar p_D +\bar p_N]-\frac{\bar p_N}{1-\bar p_D+\bar p_N}\Big)-\frac{1}{2}\Big(\log [1-\bar q_D +\bar q_N]-\frac{\bar q_N}{1-\bar q_D+\bar q_N}\Big).
\end{split}
\end{equation}
On the saddle point, we find the following self-consistency equations for the conjugates order parameters:
\begin{equation}
\bar q_D= \alpha\beta^2 p_D,\quad
\bar p_D= \beta^2 q_D,\quad
\bar q_N=\alpha\beta^2 p_N,\quad
\bar p_N=\beta^2 q_N.
\end{equation}
By inserting these equations in the formula \eqref{eq:frep}, one precisely finds the expression \eqref{eq:freestoc} for the quenched free energy.

\subsection{Self-consistency equations and extremal free energy}
In order to find the explicit expression for the quenched free energy as a function of the system parameters $(\alpha, \beta)$ the expression \eqref{eq:freestoc} must be extremized with respect to the order parameters $q_{D,N}$, $p_{D,N}$. To this goal it is convenient to rewrite the quenched free energy in terms of the two new order parameters
\begin{equation}
\begin{split}
\Delta_q &= q_D - q_N,\\
\Delta_p &= p_D - p_N,
\end{split}
\end{equation}
{\it i.e.}, the difference between the diagonal and non-diagonal overlaps. In this way, the quenched free energy can be rewritten as
\begin{equation} \label{eq:freee}
\begin{split}
f(\alpha,\beta)=&-\frac{\alpha}{2}\log(1-\beta^2\Delta_q)-\frac{1}{2}\log(1-\alpha \beta^2\Delta_p)\\&+\frac{\alpha\beta^2  p_N}{2(1-\alpha\beta^2\Delta_p)}+\frac{\alpha\beta^2  q_N}{2 (1-\beta^2\Delta_q)}-\frac{\alpha\beta^2}{2}( p_N \Delta_q+ q_N \Delta_p+\Delta_q \Delta_p).
\end{split}
\end{equation}
By imposing the extremality condition on the quenched free energy, for these order parameters we find algebraic self-consistency equations, namely
\begin{equation}
\Delta_p = \frac{1}{1-\beta^2 \Delta_q},\quad \Delta_q = \frac{1}{1-\alpha\beta^2 \Delta_p},\quad q_N = \alpha\beta^2 p_N \Delta_q^2,\quad p_N=\beta^2 q_N \Delta_p^2.
\end{equation}
By solving this algebraic system, beyond the trivial solution $p_N=q_N=0$ for non-diagonal overlaps, we get\footnote{There is also another solution for the self-consistency equations system. However, in the $\beta\rightarrow0$ limit, the diagonal overlaps diverge for such a solution. This is not a consistent behaviour since, in this limit, the two variable sets decouple in the partition function, so that the average of the diagonal overlaps are simply given by the Gaussian integrals of $z^2$ and $\varphi^2$, which of course are equal to $1$. On the other hand, the solution \eqref{eq:scsol} has the correct scaling behaviour.}
\begin{eqnarray}\label{eq:scsol}
\begin{split}
q_D \equiv \Delta_q &=\frac{1+(1-\alpha)\beta^2-\sqrt{(1+\beta^2-\alpha\beta^2)^2-4\beta^2}}{2\alpha\beta^2}	,\\
p_D \equiv \Delta_p &=\frac{1-(1-\alpha)\beta^2-\sqrt{(1+\beta^2-\alpha\beta^2)^2-4\beta^2}}{2\alpha\beta^2}.
\end{split}
\end{eqnarray}
By plugging these expressions in the quenched free energy (\ref{eq:freee}), the final result reads as
\begin{equation}\label{eq:extfe}
\begin{split}
f(\alpha,\beta)&=\frac{\alpha  \beta ^2+\beta ^2-1+\sqrt{(\alpha -1)^2 \beta ^4-2 (\alpha +1) \beta ^2+1}}{4 \beta ^2}\\&- \frac\alpha 2 \log \Big(\frac{1}{2} [(\alpha -1) \beta ^2+\sqrt{(\alpha -1)^2 \beta ^4-2 (\alpha +1) \beta ^2+1}+1]\Big)\\&-\frac12 \log \Big(\frac{1}{2} [(1-\alpha ) \beta ^2+\sqrt{(\alpha -1)^2 \beta ^4-2 (\alpha +1) \beta ^2+1}+1]\Big),
\end{split}
\end{equation}
namely, the r.h.s. of eq. \eqref{eq:extfe} is the explicit expression of  the extremal quenched free energy in the replica symmetric regime for the bipartite Gaussian spin-glass (\ref{model}).

\section{Bi-partite spin glasses and the Marchenko-Pastur law}
Random matrix theory \cite{Mehta,Tao,Vulpiani} raised around the early decades of the past century in the context of multivariate statistics by pioneers as Wishart \cite{Wishart} and Hsu \cite{Hsu} and probably became a discipline by its own since, in the fifties, Wigner systematically developed its foundations \cite{Wigner}. The crucial point is that, for several problems, in the asymptotic regime ({\it i.e.}, in the thermodynamic limit), random matrices exhibit {\em universality}, namely results become independent of the details of the original probability distributions generating the entries. 
Remarkably, {\em universality} is also a well identified property of spin glasses \cite{Univ1,Univ2}.
\newline
For instance, the eigenvalue distribution of a real symmetric matrix, whose random entries are independently distributed with equal densities, converges to Wigner's semicircle law regardless of the details of the underlying entry densities.  
\newline
Another fundamental distribution is the one named after Marchenko and Pastur, who gave the limiting distribution of eigenvalues of Wishart matrices. More precisely, the main results concerning the Marchenko-Pastur distribution can be summarized as follows. Let $P$ be a sequence of integers such that $P = \alpha N$ with $\alpha \in \mathbb{R}^+$ and consider an $N \times P$ matrix $\boldsymbol{\xi}$ whose entries are i.i.d. standard Gaussian variables, {\it i.e.}, $\xi_{i}^{\mu} \sim \mathcal{N}[0,1]$ for any $i=1,..,N$ and $\mu=1,...,P$. Normalize this matrix to get $\boldsymbol{\hat{\xi}}:=\boldsymbol{\xi}/\sqrt{N}$ and construct $\mathbf{C} = \boldsymbol{\hat{\xi}} \boldsymbol{\hat{\xi}}^T$ that is symmetric (hence has real eingenvalues $\gamma_1,...,\gamma_N$) with spectral distribution referred to as $\rho(\gamma)$. Then, the Marchenko-Pastur theorem guarantees that $\rho(\gamma)$ (weakly) converges to the distribution, defined for $\gamma \in [\gamma_-, \gamma_+]$ 
\begin{eqnarray}\label{MPtheorem}
\rho(\gamma) &=& \frac{\sqrt{(\gamma-\gamma_-)(\gamma-\gamma_+)}}{2 \pi \gamma} \bold{1}_{\gamma \in  [\gamma_-, \gamma_+]},\\
\gamma_- &=& (1 - \sqrt{\alpha})^2,  \ \ \ \gamma_+ = (1 + \sqrt{\alpha})^2,
\end{eqnarray}
that is called the Marchenko-Pastur distribution.

\subsection{Another link between Random Matrices and Statistical Mechanics}
An interesting application of the solution of the analog bipartite spin-glass lies exactly in random matrix theory, as we are going to show in this subsection. The partition function under consideration (see eq. (\ref{eq:part})) is of trivial solution via standard Gaussian integration. In fact, by direct calculations, one gets
\begin{equation}\label{party-zion}
Z_{N,P}(\beta|\xi) = {\det} ^{-1/2}(1-\beta^2 \mathbf{C}),
\end{equation}
where we defined the correlation matrix of the couplings as
\begin{equation}\label{corr-matr}
C_{\mu\nu}=\frac1N \sum_{i=1}^N \xi^\mu _i \xi^\nu_i.
\end{equation}
It is worth stressing that $\mathbf{C}$ is known as {\it Wishart matrix} in the random matrix theory and, also, it is exactly the {\it Hebbian kernel} adopted in neural networks \cite{Amit}. 
\newline
When $\beta^2 \lVert \mathbf{C} \rVert<1$, we expand the determinant as
\begin{equation}
\det (1-\beta^2 \mathbf{C})=\exp \left(-\sum_{k=1}^\infty \frac{\text{Tr }\beta^2\mathbf{ \mathbf{C}}^n}{n} \right).
\end{equation}
Taking the logarithm of the partition function (as coded at the r.h.s. of eq. (\ref{party-zion})) results in the following relation
\begin{equation}
\log Z_{N,P}(\beta|\xi) =\frac12 \sum_{k=1}^\infty \frac{\beta^{2k} \text{Tr } \mathbf{C}^k}{k}.
\end{equation}
Finally, taking the average over the weight realizations and taking the $N\rightarrow \infty $ limit after multiplying by $1/N$, we have
\begin{equation}
f(\alpha,\beta)=\frac12\sum_{k=1}^\infty\frac{\beta^{2k}}{k} c_k,
\end{equation}
where
\begin{equation}
c_k =\lim_{N\rightarrow\infty} \frac1N \mathbb E_\xi \text{Tr } \mathbf{C}^k.
\end{equation}
In other words, the quenched free energy of the model (\ref{model}) is the generating function of the moments of the correlation matrix of the weights ({\it i.e.} of the {\em patterns} in a neural network jargon). Therefore, we can now generate them with the elementary formula
\begin{equation}\label{eq:momentafree}
\lim_{N\rightarrow\infty} \frac1N \mathbb E_\xi \text{Tr } \mathbf{C}^n= \frac{2 n  }{n!}\Big(\frac{d}{d\beta^2}\Big)^n f\Big\vert_{\beta=0}.
\end{equation}
From random matrix theory, we know that the momenta of Wishart matrix capture all the information about its spectrum. Indeed,\footnote{We stress that the $\alpha$ factor is here due to the normalization constant $N^{-1}$ instead of $P^{-1}$.}
\begin{equation}
\lim_{N\rightarrow\infty}\frac1N \mathbb E_\xi \,\text{Tr}\, C^n=\alpha \mathbb E_\xi[ \gamma^n] =\alpha \int d\gamma \gamma^n \rho (\gamma),
\end{equation}
where $\gamma$ is the generic eigenvalue of the correlation matrix and $\rho$ is the relative spectral probability distribution. The integral in the last line is the $n$-th eigenvalue distribution moment. The generating function is trivial in this case and corresponds to the so-called Stieltjes transformation of $\rho$ given by
\begin{equation}\label{GF2}
\int d\gamma \frac{\rho(\gamma)}{z-\gamma},
\end{equation}
where $z$ is complex variable away from the real axis. In fact, with a Laurent expansion of the generating functional for $\vert z\vert \rightarrow \infty$, one precisely generates the moment of the spectral distribution. Then, in order to compare both the generating functions, the one stemming from disordered statistical mechanics (see eq. \ref{eq:extfe}) and the standard one from random matrix theory (see eq. \ref{GF2}), we obtain the following novel bridge
\begin{equation}
\begin{split}
\alpha \int d\gamma \frac{\rho(\gamma)}{z-\gamma}&= \frac\alpha z+\frac\alpha z\sum_{n=1}^\infty z^{-n}  \int d\gamma \gamma^n \rho (\gamma)=\frac\alpha z+\frac{2}{z}\sum_{n=1}^\infty \frac{z^{-n}}{(n-1)!}\Big(\frac{d}{d\beta^2}\Big)^n f\Big\vert_{\beta=0}=\\
&=\frac\alpha z+\frac{2}{z^2}\Big[\frac{d}{d\beta^2}\sum_{n=1}^\infty \frac{z^{-(n-1)}}{(n-1)!}\Big(\frac{d}{d\beta^2}\Big)^{n-1} f\Big]_{\beta=0}.
\end{split}
\end{equation}
The series in brackets can be trivially re-summed in $\exp(z^{-1}\partial_{\beta^2})f$, generating a (complex) translation of $1/z$ in the variable $\beta^2$, so that
\begin{equation}
\alpha \int d\gamma \frac{\rho(\gamma)}{z-\gamma}=\frac\alpha z+\frac{2}{z^2}\Big[\frac{d}{d\beta^2}f(\beta^2+1/z)\Big]_{\beta=0}.
\end{equation}
This equality directly implies that
\begin{equation}
\int d\gamma \frac{\rho(\gamma)}{z-\gamma}=\frac1z+\frac{2}{\alpha z^2}f'\left(\frac1z\right).
\end{equation}
This is a nice exact result stating that the Stieltjes transform of the spectral density is captured by the quenched free energy corresponding to the replica symmetric solution of the bi-partite analog spin-glass.
\newline
Further, the knowledge of the exact form of the quenched free energy allows us to reconstruct the spectral density by using the Stieltjes-Perron inversion formula: to do this, we stress that the derivative of the quenched free energy (once computed in $1/z$) is
\begin{equation}
f'\left(\frac1z\right)=\frac{1}{z}+\frac{2}{z[z-(\alpha+1) +\sqrt{(\alpha -1)^2+z^2-2 (\alpha +1) z}]}.
\end{equation}
Because of the presence of the square root in the denominator, this function has a branch cut on the segment connecting the points
\begin{equation}
\gamma_{\pm}= (1\pm \sqrt \alpha)^2,
\end{equation}
on the real axis.\footnote{Note that $\gamma_+$ shares the same expression in $\alpha$ as the critical temperature for ergodicity breaking for the Hopfield model of neural networks \cite{Amit}.} Then, by computing the jump across the whole $\mathfrak R (z)$ and finally applying the Stieltjes-Perron formula, we re-obtain the well-known distribution
\begin{equation}\label{eq:marchenko}
\rho (\gamma)= \frac{\sqrt{(\gamma-\gamma_-)(\gamma_+ -\gamma)}}{2\pi \alpha \gamma} \bold{1}_{\gamma \in  [\gamma_-, \gamma_+]},
\end{equation}
where $\bold{1}_{\gamma \in  [\gamma_-, \gamma_+]}$ is the characteristic function of the interval $[\gamma_{-},\gamma_{+}]$: this is the well-known Marcenko-Pastur law (see eq. \ref{MPtheorem}).
\newline
There is a certain similarity of this setting with Hebbian neural networks:
\begin{itemize}
\item even in the original formulation of the Marchenko-Pastur law derivation, the {\em high storage} behavior \cite{Amit} is required \cite{Mehta}, namely we must face $\alpha \sim P/N \in (0,1])$ in the infinite volume limit.

\item upon marginalization of the partition function (\ref{party-zion}) over the dynamical variables $\bold{\varphi}$ we approach
    $$
    Z_{N,P}(\beta|\xi)= \int d\mu (\bz,\bfi)\exp\Big(\frac{\beta}{\sqrt{N}}\sum_{\mu, i}^{P, N}\xi^\mu _i z_\mu \varphi_i\Big)=\int d\mu(\bz)
    \exp\left[ \frac{\beta^2}{2N}\sum_{\mu,\nu}^{P,P} \left(\sum_{i}^{N}\xi_i^{\mu}\xi_i^{\nu} \right)z_{\mu} z_{\mu} \right]
    $$
    that is the partition function of the spherical Hopfield model \cite{spherical,gaussian-spinglass,Genovese-Tantari}.

\item there is a purely probabilistic perspective by which it is immediate to recognize that the Marchenko-Pastur law is deeply related to bipartite spin glasses: by a Jaynes inferential perspective \cite{Jaynes}, thus looking at entropy maximization as a statistical requirement \cite{Bialek}, it is clear that two-body interactions in statistical mechanics are related to correlations in statistical inference. As correlations can be positive as well as negative it is obvious why we do need spin-glass architectures rather than {\it e.g.} ferromagnetic ones.
    \newline
    Also, it is clear why we actually need a bipartite structure: if approaching the proof of the Marchenko-Pastur law coded by Eqs.  (\ref{MPtheorem}), for instance with the momenta method, it is crystal clear that we must face expectations whose structure is $\propto \sum_{i_1,...,i_k}\sum_{j_1,...,j_k}\mathbb{E}_{\xi}\xi_{i_1}^{\mu_1}\cdot \xi_{i_2}^{\mu_1} \cdot \xi_{i_2}^{\mu_2} ...\xi_{i_k}^{\mu_k}$, where $i_k \in (1,...,N)$ and $j_k \in (1,...,P)$. Now, as the entries are independent, each factor in the product $\mathcal{P}\equiv \xi_{i_1}^{\mu_1}\cdot \xi_{i_2}^{\mu_1} \cdot \xi_{i_2}^{\mu_2} ...\xi_{i_k}^{\mu_k}$ must appear twice in order for the expectation $E_{\xi}(P) \neq 0$ (as the Gaussians are centered and with unitary variance): we can thus think at the two summations as labelling a bi-partite undirected graph whose layers have sizes, respectively, of $N$ and $P$ variables (namely the analog spin-glass under investigation).
\end{itemize}

\subsection{Numerical comparisons between the two routes}
\begin{figure}[b!]
	\centering
	\includegraphics[scale=0.5]{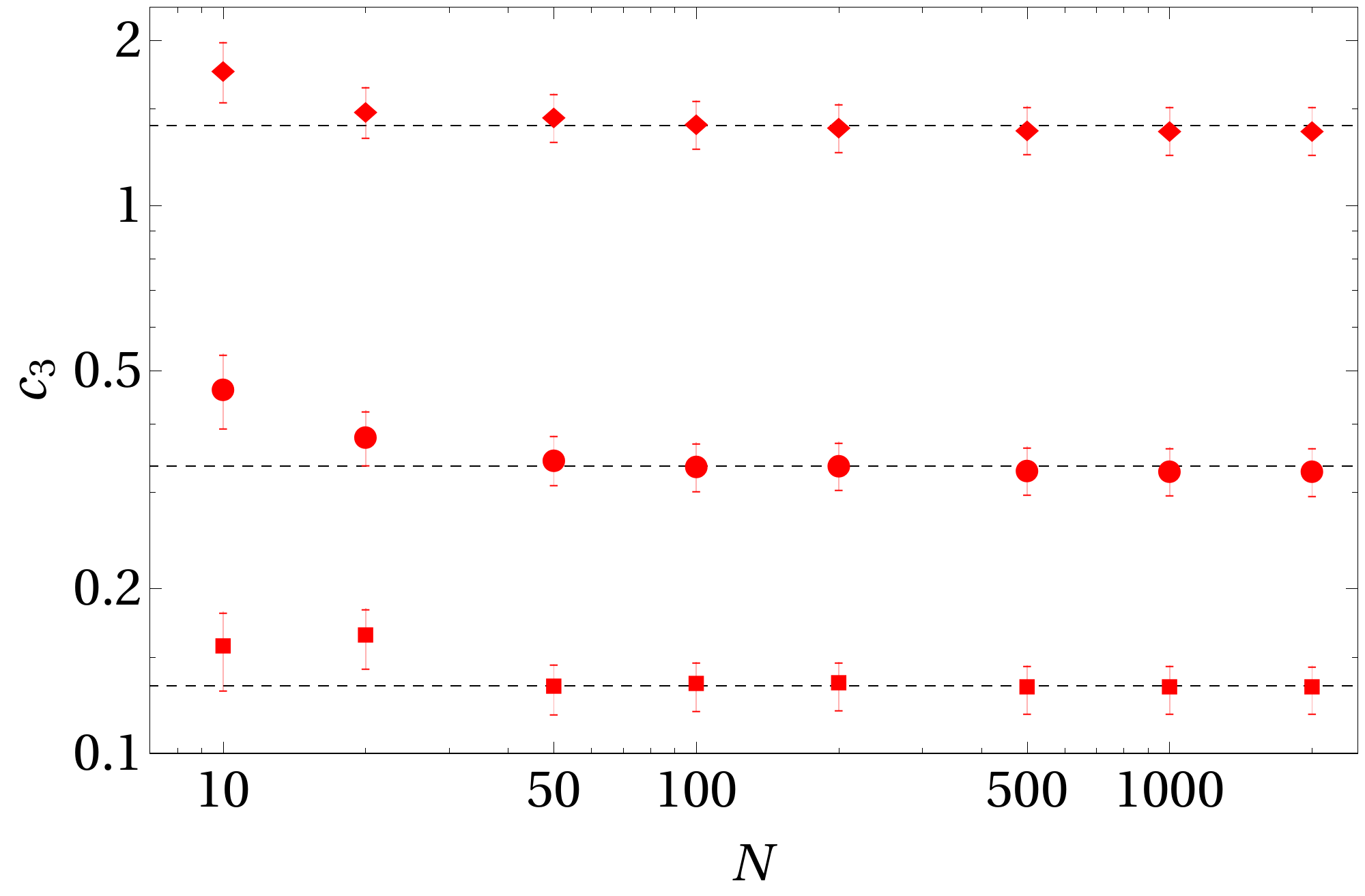}		
	\caption{{\bfseries Finite size scaling for $c_3$.} The plot shows the dependence on $N$ of the third moment of the correlation matrix. The data are obtained by the average over $50$ different realizations for each $N$ and $\alpha=0.1$ (red squares), $0.2$ (red circles) and $0.5$ (red diamonds). The dashed black lines are the relative $N\rightarrow \infty$ values of $c_3$ coming from the finite-size scaling analysis.}\label{fig:c1fss}
\end{figure}
We now compare the theoretical outcomes predicted by the analytical expression of the quenched free energy (Disordered Statistical Mechanics) with the numerical estimations of the momenta of the correlation function (Random Matrix Theory).\par\bigskip
Concerning the latter, on the numerical side, we consider 50 different realizations of the couplings $\boldsymbol\xi$ for $N=10,20,50,100,\dots ,2000$ and $\alpha =0.1, 0.2,\dots,0.9$, and then we compute the powers of correlation matrix (we restricted the inspection to the lowest orders $n=1,2,3,4,5$) and finally we average over the samples. For increasing $N$, at fixed $\alpha$, the momenta settle on their ``thermodynamic'' values. We therefore perform a finite-size scaling analysis by fitting numerical data for sufficiently large $N$: an example of this analysis is reported in Fig. \ref{fig:c1fss} for the momenta $c_3$ for $\alpha=0.1,0.2,0.5$.
\par\medskip
\begin{figure}[t!]
	\centering
	\includegraphics[scale=0.5]{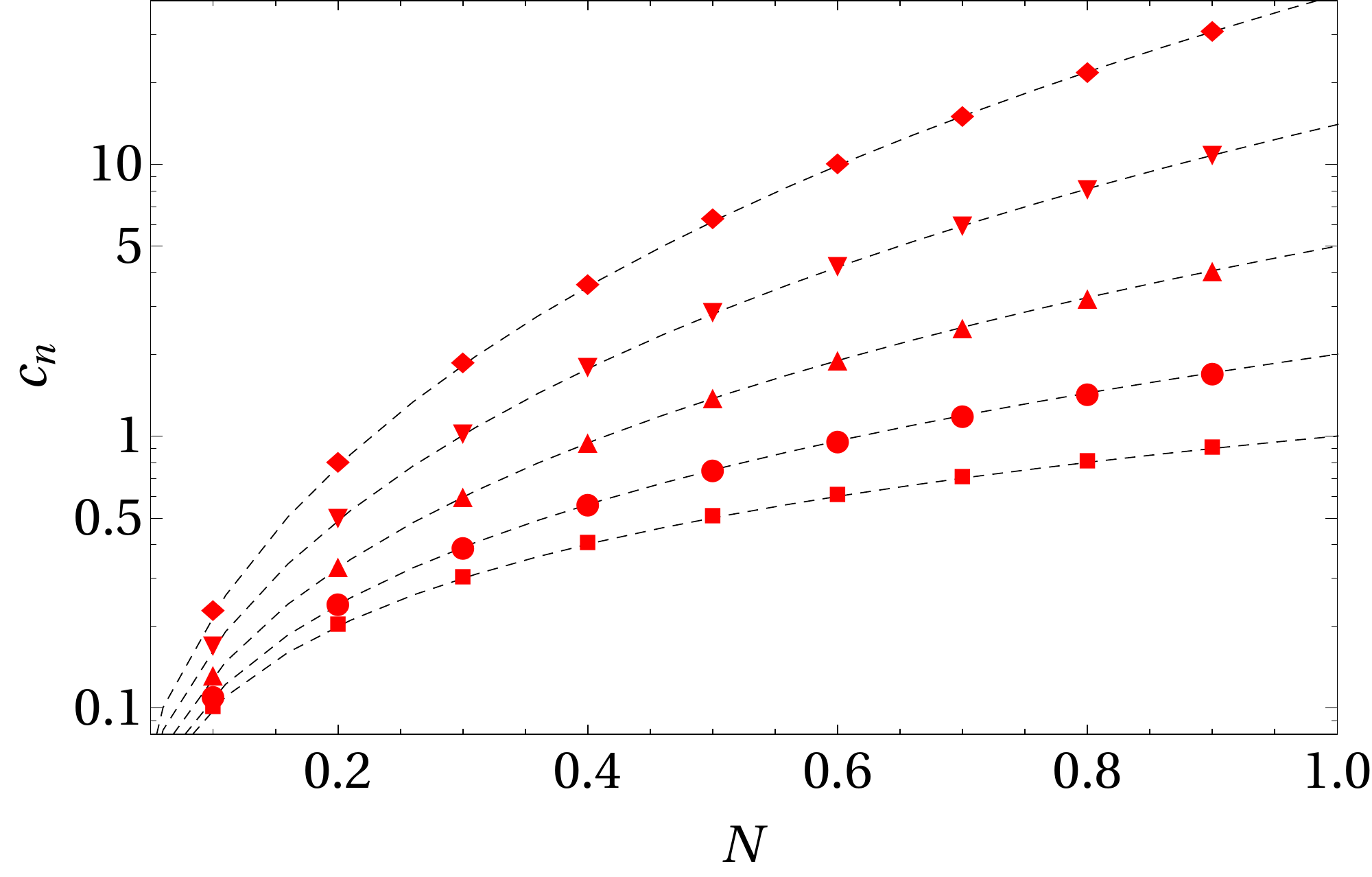}
	\caption{{\bfseries Comparison between theoretical predictions and numerical results.} The plot shows the comparison between the numerical results (obtained with the finite-size scaling analysis for the correlation matrix momenta) and the theoretical prediction derived by the generating function \eqref{eq:extfe}, which are explicitly collected in \eqref{eq:theomomenta}. The error bars for numerical data are too small to be visualized in the plot (because of the logarithmic scale).}\label{fig:comp}
\end{figure}
\begin{figure}[b!]
	\centering
	\begin{minipage}[c]{.49\textwidth}
		\centering
		\includegraphics[width=\textwidth]{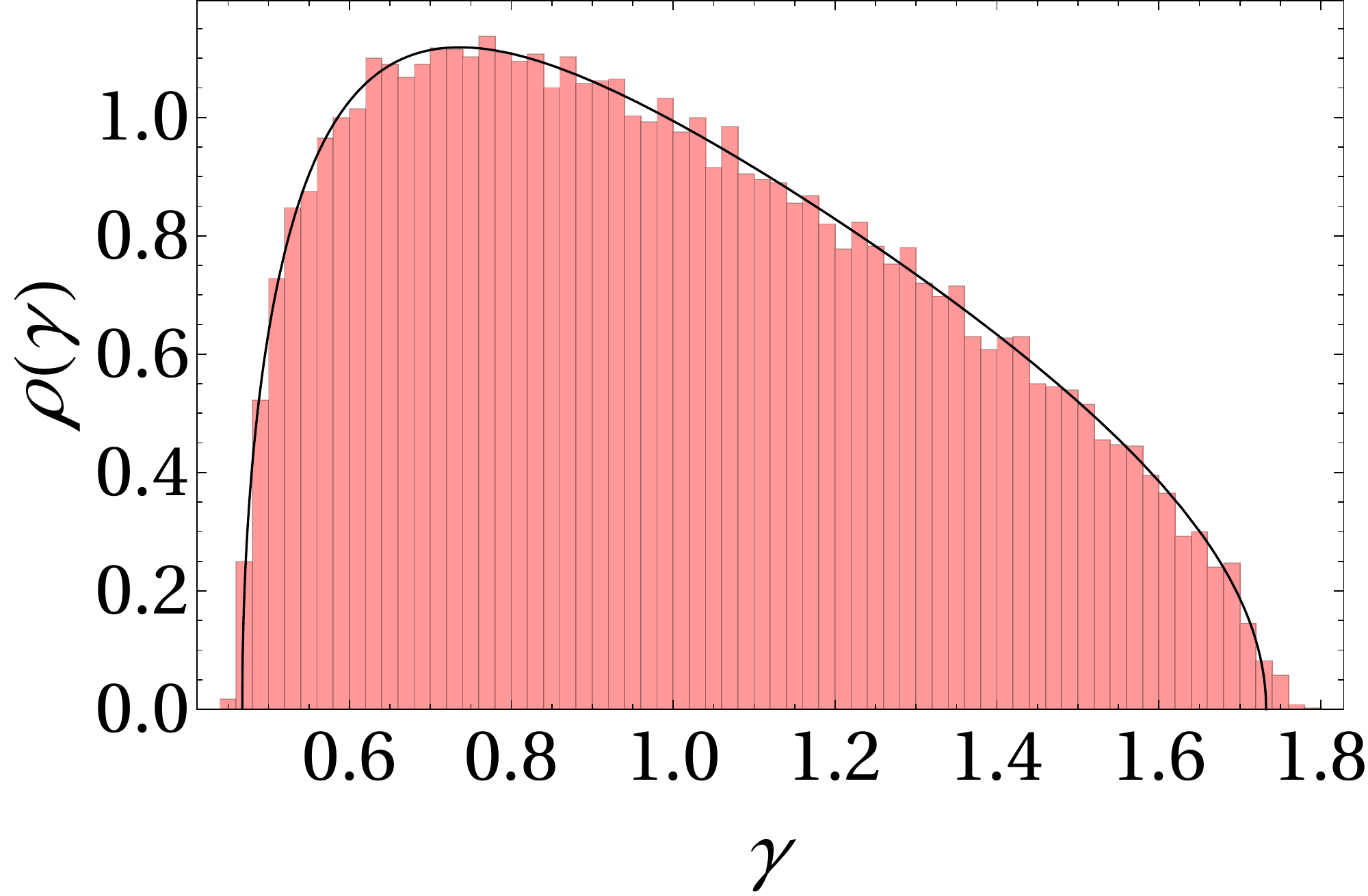}
	\end{minipage}
	\begin{minipage}[c]{.49\textwidth}
		\centering
		\includegraphics[width=\textwidth]{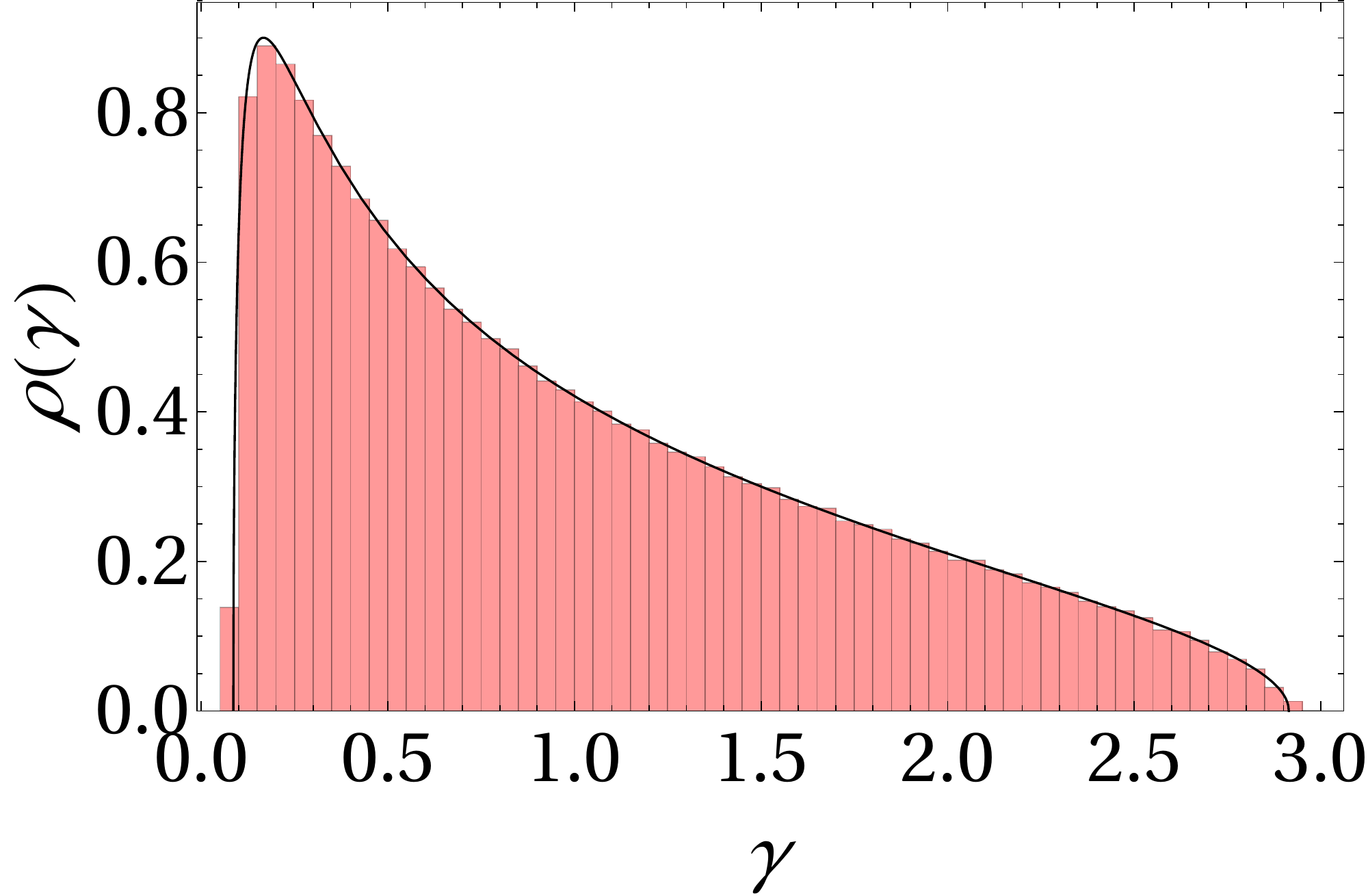}
	\end{minipage}
	\caption{{\bfseries Spectral density versus frequency distributions.} The plots show the eigenvalues histograms versus the spectral probability distribution \eqref{eq:marchenko}. Numerical data are obtained from a sample of 200 correlation matrices with $N=1000$ and $\alpha=0.1$ (left plot) and $\alpha=0.5$ (right plot).}\label{fig:march}
\end{figure}
From the analytical side, by using both \eqref{eq:momentafree} and \eqref{eq:extfe}, we can easily determine the theoretical predictions for the correlation matrix momenta. For $n=1,\dots, 5$, we easily find
\begin{equation}\label{eq:theomomenta}
\begin{split}
c_1&=\alpha,\\
c_2&=\alpha(\alpha+1),\\
c_3&=\alpha  \left(\alpha ^2+3 \alpha +1\right),\\
c_4&=\alpha  (\alpha +1) \left(\alpha ^2+5 \alpha +1\right),\\
c_5&=\alpha  \left(\alpha ^4+10 \alpha ^3+20 \alpha ^2+10 \alpha +1\right).
\end{split}
\end{equation}
The comparison between the numerical estimations and the theoretical predictions is reported in Fig. \ref{fig:comp}, showing a perfect agreement between the numerical data and the analytical solutions from the quenched free energy of the analog bi-partite spin-glass.
\par\bigskip
For the sake of completeness, we compare also the Marchenko-Pastur law \eqref{eq:marchenko} derived in our setup with numerical simulations. To do this, we considered a sample of $200$ correlation matrices with $N=1000$ and $\alpha=0.1,0.5$. Then, we computed the eigenvalues of each matrix in the sample and realize the histogram. The latter is compared with the spectral probability distribution according to Marchenko-Pastur law. Also in this case, it emerges a perfect agreement between numerical data and the analytical solution, as it emerges from Fig \ref{fig:march}.

\section{Conclusions}

Bipartite spin-glasses \cite{bipartiti} have recently attracted much attention (see {\it e.g.} \cite{Auffinger-bip,Barra-Multispecies,Gavin-bip,Monasson} and references therein)  as they constitute the natural mathematical scaffold for Boltzmann machines \cite{Hinton1,HintonLast}, the latter being the building blocks of powerful inferential architectures ({\it e.g.} the so-called {\em deep Boltzmann machines} \cite{Hugo}) in Machine Learning.
\newline
Bipartite spin-glasses can be equipped with binary or real valued spins and/or couplings and, as extensively revised in \cite{Barra-RBMsPriors1,Barra-RBMsPriors2}, a ``quasi-universal'' behavior emerges in these models ({\it i.e.} {\em solely} the retrieval properties are strongly sensible to the nature of these variables but the resulting structure of the quenched noise -coded in the overlaps- is unaffected by the details of the priors), a property share by random matrices, whose study lies in Random Matrix Theory.
\newline
In this work we focused on the fully analog bipartite spin-glass, namely a two-layer network whose layers (or parties) share standard Gaussian spins and also the (quenched) couplings among the spins are drawn
from standard i.i.d. Gaussian distributions. The main route, in the disordered statistical mechanical analysis of the model, is to look for an explicit expression of quenched free energy related to the cost function defining the model, possibly in terms of the natural parameters of the theory, that -in the present case- turn out to be the self and two-replica overlaps, for each layer.
\newline
Confining our investigation to a replica symmetric scenario (that is however expected to be correct for these analog models \cite{gaussian-spinglass,BenArous}), we obtained an explicit expression for the quenched free energy by using both the interpolation technique \cite{Barra-JSP2010,Guerra-Broken} and the replica-trick \cite{MPV}, finding overall perfect agreement among the outcomes of the two computations.
\newline
A remarkable property of this analog model is that it is entirely approachable even without the knowledge of statistical mechanical techniques (as, being the model entirely Gaussian it is mathematically tractable) and a direct calculation shows that the quenched free energy of this model is the generating functional for the momenta of the related Willshaw matrix (namely the correlation matrix among the weights, that -not by chance- turns out to be Hebbian \cite{Amit,BarraEquivalenceRBMeAHN,Coolen}): this novel bridge allows drawing a number of conclusions. In particular, we have shown here how -{\em reversing the perspective}- we can obtain information in Random Matrix Theory by using Statistical Mechanics: by applying the Stieltjes-Perron inversion formula on the quenched free energy of the analog bi-partite spin-glass  the spectral density of the Wishart matrix (or Hebbian kernel) can be obtained in terms of the celebrated Marchenko-Pastur distribution.
\newline
These analytical findings have been also supported by extensive numerical simulations finding overall full agreement.


\section{Acknowledgments}
The Authors are grateful to Alessia Annibale and Reimer K\"{u}hn for useful discussions. The Authors acknowledge partial financial fundings by MIUR via {\em FFABR2018-(Barra)} and by INFN.

\end{document}